\documentclass[conference]{IEEEtran}
\IEEEoverridecommandlockouts

\usepackage{graphicx}
\usepackage{xcolor}
\newcommand{\linebreakand}{%
  \end{@IEEEauthorhalign}
  \hfill\mbox{}\par
  \mbox{}\hfill\begin{@IEEEauthorhalign}
}    
\usepackage{amsmath}
\usepackage{amssymb}
\usepackage{cite}
\usepackage{bbm}
\usepackage{algorithm} 
\usepackage{algpseudocode}
\usepackage{caption}
\usepackage{steinmetz}
\usepackage{subcaption}
\usepackage{amsthm}
\usepackage[english]{babel}

\newcounter{relctr} 
\everydisplay\expandafter{\the\everydisplay\setcounter{relctr}{0}} 
\DeclareMathOperator*{\argmax}{argmax} 
\DeclareMathOperator*{\minimize}{minimize} 
\DeclareMathOperator*{\maximize}{maximize} 
\usepackage[figurename=Fig.]{caption}

\AtBeginDocument{} 

\ifCLASSINFOpdf
\else
\fi

\usepackage[normalem]{ulem}

\begin{document}
\vspace{-4mm}
\title{Deep Reinforcement Learning for Multi-User RF Charging with Non-linear Energy Harvesters \\
\thanks{This work is partially supported by the Research Council of Finland (Grants 348515 and 346208 (6G Flagship)); by the European Commission through the Horizon Europe/JU SNS project Hexa-X-II (Grant no. 101095759); by the Finnish-American Research \& Innovation Accelerator; and by the Villum Investigator Grant "WATER” from the Velux Foundation, Denmark.}
}


\author{
{ Amirhossein~Azarbahram$^{*}$, Onel~L.~A.~L\'{o}pez$^{*}$, Petar~Popovski$^{\dagger}$, Shashi~Raj~Pandey$^{\dagger}$, and Matti Latva-aho$^{*}$}
\vspace{1mm}
\\

$^{*}$Centre for Wireless Communications (CWC), University of Oulu, Finland\\
$^{\dagger}$Department of Electronic Systems, Aalborg University, Denmark \\

\normalsize Emails: \{amirhossein.azarbahram, onel.alcarazlopez\}@oulu.fi, \{petarp, srp\}@es.aau.dk, and matti.latva-aho@oulu.fi
\vspace{-4mm}


}

\maketitle

\begin{abstract}
Radio frequency (RF) wireless power transfer (WPT) is a promising technology for sustainable support of massive Internet of Things (IoT). However, RF-WPT systems are characterized by low efficiency due to channel attenuation, which can be mitigated by precoders that adjust the transmission directivity. This work considers 
a multi-antenna RF-WPT system with multiple non-linear energy harvesting (EH) nodes with energy demands changing over discrete time slots. This leads to the charging scheduling problem, which involves choosing the precoders at each slot to minimize the total energy consumption and meet the EH requirements. We model the problem as a Markov decision process and propose a solution relying on a low-complexity beamforming and deep deterministic policy gradient (DDPG). The results show that the proposed beamforming achieves near-optimal performance with low computational complexity, and the DDPG-based approach converges with the number of episodes and reduces the system's power consumption, while the outage probability and the power consumption increase with the number of devices.
\end{abstract}

\begin{IEEEkeywords}
Radio frequency wireless charging, energy beamforming, near-field channels, charging scheduling.
\end{IEEEkeywords}

\vspace{-2mm}

\section{Introduction}

\IEEEPARstart{T}{he} massive growth of the Internet of Things (IoT) networks threatens to skyrocket the maintenance cost of these systems, especially those related to battery replacement and waste. This calls for practical alternatives such as relying on radio frequency (RF) wireless power transfer (WPT) technology. RF-WPT can provide wireless charging capabilities to prevent IoT devices from battery depletion and increase their lifespan \cite{3gppamb, ZEDHEXA}. Also, it can potentially charge multiple devices over large distances and use the same infrastructure as wireless communications by utilizing the broadcast nature of the wireless channel. However, the inherently low efficiency of RF-WPT systems is still an open challenge, demanding significant attention \cite{intro3, lópez2023highpower}. One of the main inefficiency sources is channel attenuation, which can be mitigated by energy beamforming (EB). Furthermore, there is a need to properly allocate charging resources in a multi-node IoT system to avoid interruptions during the system operation time.

EB techniques can compensate for the channel losses by focusing the energy beams on the devices \cite{intro3}. For example, the authors in \cite{OnelRadioStripes} propose EB designs to power multiple devices in a radio stripe system, while the deployment problem of the radio stripes using maximum ratio transmission beamformers is investigated in \cite{azarbahram2023radio}. In \cite{azarbahram2023EB_DMA,DMAWPT}, the authors propose EB approaches for near-field RF-WPT with dynamic metasurface antennas as the transmitter. Moreover, a low-complexity EB design is proposed in \cite{onellowcomp} to power single-antenna devices. 

There are some works in the literature on improving the performance of energy harvesting (EH)-assisted wireless systems over a time horizon. For example, the authors in \cite{RL_schedul_TGC2019} consider a wireless-powered sensor network and utilize deep reinforcement learning (DRL) to select a node to charge and its corresponding allocated power at each time slot to minimize the packet loss caused by insufficient energy. In \cite{RL_IOD}, deep deterministic policy gradient (DDPG) is used to tune the transmit power to minimize long-term energy consumption. The authors in \cite{DRL_EHWN_TGC2021} utilize DDPG to improve the energy efficiency in a heterogeneous network by tuning transmit/harvest strategies for macro or small cells at each time slot. Interestingly, DRL is used in \cite{RL_adaptive_ResourceWPCN_ICL2020} to split the channel resources between EH and information transmission using DRL aiming to minimize the outage probability of the information transmission phase. Moreover, in \cite{RIS_UAV_WPT_WIT_RL_JSAC2022}, the authors consider a WPCN with reflective intelligent surfaces and unmanned aerial vehicles and jointly optimize the transmit power, trajectory, and phase shifts using DRL.

Although there are many works considering DRL-based approaches for intelligent charging, no work has yet addressed the joint beamforming and charging scheduling problem in RF-WPT systems. Herein, we aim to precisely fill this research gap by considering a multi-antenna RF-WPT system with multiple non-linear EH devices with energy requirements at each time slot. Our main contributions are: i) we formulate the charging scheduling problem, which consists of choosing precoders at each time slot aiming to minimize the average power consumption and meet the EH requirements of the nodes; ii) we propose a solution relying on DDPG and a low-complexity beamforming design; iii) the results evince that the proposed beamforming achieves near-optimal performance with much less complexity, while the power consumption decreases and the reward function converges to a suboptimal solution with the number of episodes in the DDPG algorithm. 

\textbf{Structure:} Section~\ref{SEC:system} introduces the system model and the problem formulation. The proposed optimization framework is discussed in Section~\ref{SEC:Optimization}, while Section~\ref{SEC:numerical} provides the numerical analysis, and Section~\ref{SEC:conclusion} concludes the paper. \textbf{Notations:} Bold-face and non-Bold-face characters refer to vectors and scalars, respectively, $(\cdot)^H$ denotes the Hermitian operator, $\Re\{\cdot\}$ is the real operator, $\mathbbm{1}(\cdot)$ is the indicator function, while $\mathbf{0}$ and $\mathbf{1}$ refer to all-zeros and all-ones vector, respectively.

\section{System Model}\label{SEC:system}

The system operates in $T$ time slots, where a fully-digital antenna array with $N$ elements charges $K$ EH nodes with $d_{i,k}$ being the energy demand of the $k$th device in the $i$th slot, which is unknown to the transmitter.

\begin{figure}[t]
    \centering
    \includegraphics[width=\columnwidth]{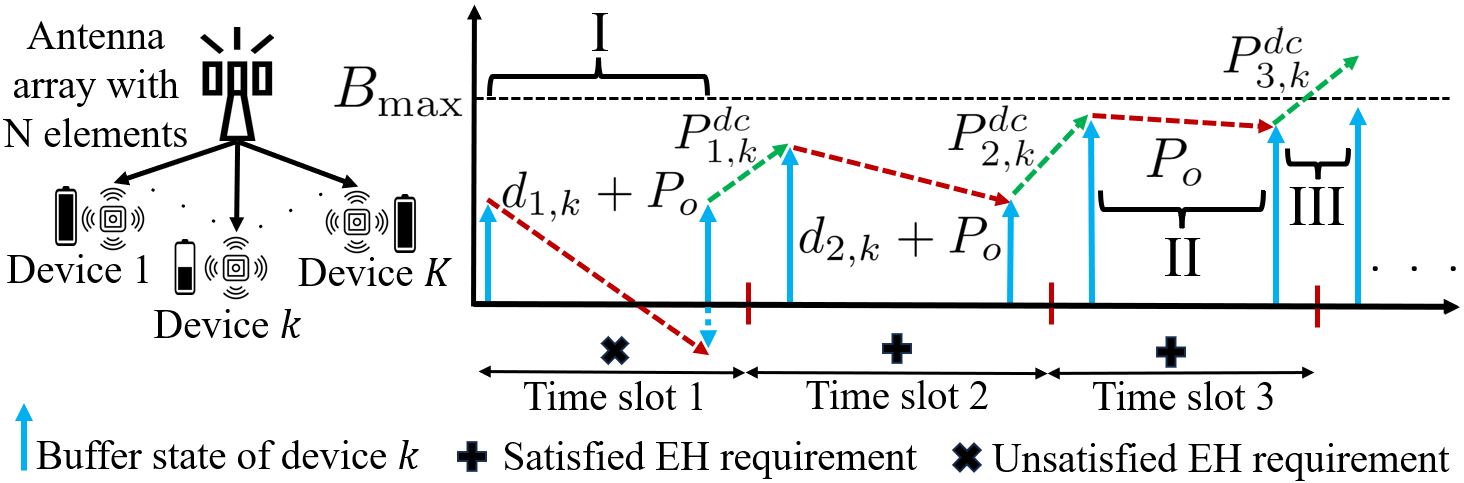}
    \caption{The abstract system (left) and the buffer model (right). Case I: there is not sufficient energy to satisfy the EH requirement; case II: the device is in idle mode; and case III: all the incident power cannot be harvested due to buffer size limitation ($B_{\max})$.}
    \label{fig:sysmod}
    \vspace{-6mm}
\end{figure}

\subsection{Signal \& Channel Model}

We consider $M \leq \min(N, K)$ energy symbols with $x_{i, m}$ being the $m$th normalized energy symbol at slot $i$ such that $\mathbb{E}\{x_{i,m}x^H_{i,n}\} = 0$, $\mathbb{E}\{x_{i,m}x^H_{i,m}\} = 1$, and $\mathbf{w}_{i,m} \in \mathbb{C}^{N \times 1}$ being the corresponding digital precoder. Thus, the transmit signal at time slot $i$ is $\mathbf{z}^{\textrm{tx}}_i = \sum_{m=1}^{M} \mathbf{w}_{i,m}x_{i,m}$. Motivated by the common use of  RF-WPT in indoor environments with line-of-sight (LOS) and near-field communication, as in smart homes, automated warehousing, and medical instruments \cite{3gppamb}, we utilize the near-field LOS channel model from~\cite{DMAWPT}. 

Let us define the Cartesian coordinate of the $l$th radiating element as $\mathbf{g}_{l}$. The channel coefficient between user $k$ and the $l$th element is given by $a_{l,k} = A_{l,k} e^{\frac{-j2\pi}{\lambda} h_{l,k}}$,
where ${\frac{2\pi}{\lambda} h_{l,k}}$ is the phase shift caused by the propagation distance, with wavelength $\lambda$, from the transmitter to the receiver and $h_{l,k} = ||\mathbf{e}_k - \mathbf{g}_{l}||$ is the Euclidean distance between the element and the user located at $\mathbf{e}_k$. Moreover,
\begin{equation}
    A_{l,k} = \sqrt{F(\Theta_{l,k})}\frac{\lambda}{4\pi h_{l,k}}
\end{equation}
is the channel gain coefficient. Here, $\Theta_{l,k} = (\theta_{l,k},\psi_{l,k})$ is the elevation-azimuth angle pair, and $F(\Theta_{l,k})$ is the corresponding radiation profile of each element, given by \cite{anetnna_radiation}
\begin{equation}
        F(\Theta_{l,k}) = \begin{cases}
        G_t\cos{(\theta_{l,k})}^{\frac{G_t}{2}-1}, & \theta_{l,k} \in [0,\pi/2],
        \\
        0, & \text{otherwise},
        \end{cases}
\end{equation}
where $G_t = 2(g+1)$ is the transmit antenna gain, and $g$ denotes the boresight gain, which depends on the antenna technology. Notably, the channel becomes $A_{k} e^{-j\psi_{l,k}}$ for far-field communications, where $A_k$ depends on the $k$th user distance and $\psi_{l,k}$ is determined by the user direction and the relative disposition of the array elements. 

\subsection{Received Signal \& EH Device Model}

We consider $\mathbf{a}_{k} = [a_{1, k}, \ldots, a_{N, k}]^T$ is the channel vector between user $k$ and the transmitter. Then, the $k$th device's received signal at time slot $i$ is $z^{\textrm{rx}}_{i,k} = \sum_{m = 1}^{M} \mathbf{a}^H_k\mathbf{w}_{i,m}x_{i,m}$, and the corresponding RF power is given by
\begin{equation}\label{eq:RCVPow}
    P^{\textrm{rf}}_{i, k} = \mathbb{E}_m\{z^{\textrm{rx}}_{i,k} {z^{\textrm{rx}}_{i,k}}^H\} = \sum_{m = 1}^{M} ||\mathbf{a}^H_k \mathbf{w}_{i,m}||^2,
\end{equation}
while $P_i^{\textrm{tx}} = \sum_{m = 1}^{M}||\mathbf{w}_{i,m}||^2$ is the $i$th slot's transmit power. In practice, the EH device is not ideal and this RF power cannot be fully utilized due to RF-to-direct current (DC) conversion inefficiencies. Specifically, there are saturation and decrements in conversion efficiency as we move toward higher input powers. Thus, we utilize the practical non-linear model presented in \cite{sigmoidmodel} with the harvested DC power of the $k$th device at slot $i$ given by
\begin{equation}\label{eq:dcpower}
    P^{\textrm{dc}}_{i, k} = \frac{\Psi_k - P^{\textrm{sat}}_k/(1 + e^{\varphi_k \omega_k})}{1 - 1/(1 + e^{\varphi_k \omega_k})},
\end{equation}
where $\Psi_k = P^{\textrm{sat}}_k/\bigl(1 + e^{-\varphi_k( P^{rf}_{i, k} - \omega_k)}\bigr)$, $P^{\textrm{sat}}_k$ is the saturation power, while $\varphi_k$ and $\omega_k$ are the parameters related to the EH circuit obtained by curve fitting on the measurement data.

Notably, each device has a buffer to store the harvested energy. The amount of available energy in the buffer of user $k$ at the beginning of time slot $i$ is written as
\begin{multline}\label{eq:bufermodel}
    b_{i, k} = \min\bigl\{b_{i-1, k} + P^{\textrm{dc}}_{i -1, k} \\- (d_{i - 1, k} + P_o)\mathbbm{1}{(b_{i-1, k} \geq d_{i - 1, k} + P_o)}, B_{\max}\bigr\},
\end{multline}
 where $B_{\max}$ is the maximum buffer size and $P_o$ is the fixed power required for keeping the EH device's circuit operational in idle mode. Equation \eqref{eq:bufermodel} indicates that devices need $P_o$ to keep functionality in the idle mode (${d}_{i,k} = 0$). Moreover, the energy demand is deducted from the buffer if there is sufficient stored energy. Without loss of generality, we have normalized the time slot duration, which leads to dealing with the average harvested power during each time slot. Fig.~\ref{fig:sysmod} further clarifies the system and buffer model.

\subsection{Problem Formulation}

The goal is to design the beamformers at each time slot to meet the EH requirements while minimizing the system's energy consumption. Thus, the optimization problem is 
\begin{subequations}\label{probnew}
\begin{align}
\label{probnewa} \minimize_{\mathbf{w}_{i, m}}  \quad & \hspace{-2mm} \frac{1}{T} \sum_{m = 1}^{M}  \sum_{i = 1}^{T} ||\mathbf{w}_{i, m}||^2 \\
\textrm{subject to} \label{probnewd} \quad & B_{\max} \geq b_{i, k} \geq d_{i, k} +  P_o, \quad \forall k, i, \\
\label{probnewe} \quad & P_i^{\textrm{tx}} \leq P_{\max}, \quad \forall i,
\end{align}
\end{subequations}
where $P_{\max}$ is the maximum transmit power at each time slot. This problem captures the charging scheduling by tuning the beamformers' direction and power toward the devices at each time slot, leading to different charging priorities and selections. Notably, \eqref{probnew} is a complex decision-making problem due to the EH non-linearity, the randomness of the EH requirements, and the correlation between the decisions in different time slots. 

\section{Optimization Framework}\label{SEC:Optimization}

The Markov decision process (MDP) is used to model a problem of such nature, which leads to efficiently handling the complexity caused by the correlation between decisions in different time slots. MDP models can be optimized to derive suboptimal decisions using DRL-based approaches. For this, MDP models rely on three main components: the action space, the state space, and the reward function. Notably, for a high-dimensional action space, one may encounter the `curse of dimensionality' when optimizing the model using DRL, i.e., the inability of the algorithm to find proper solutions \cite{DRL-survey}. This may also happen in our scenario if the action space consists of precoders ($\mathbf{w}_{i, m}$), leading to a drastic increase in the dimension of the action space, scaling with the array size. 

\subsection{MDP Model}

We proceed by defining $\alpha_{i,k}$ as a charging weight for the $k$th device and $P_i^{\textrm{tx}}$ as the transmit power of the $i$th slot. Hereby and by considering given values for $\alpha_{i, k}$ and $P_i^{\textrm{tx}}$, we formulate the beamforming problem at slot $i$ as 
\begin{subequations}\label{probnewBF}
\begin{align}
\label{probnewBFa} \maximize_{\mathbf{w}_{i, m}, \forall m} \quad &  \sum_{k = 1}^{K} \alpha_{i, k} P^{\textrm{dc}}_{i,k} \\
\textrm{subject to} \label{probnewBFb} \quad & \sum_{m = 1}^M ||\mathbf{w}_{i, m}||^2 \leq P_i^{\textrm{tx}}.
\end{align}
\end{subequations}
Now, the action space of the MDP model only consists of the charging weights, i.e., $\alpha_{i, k}$, and the transmit power, i.e., $P^{\textrm{tx}}_{i}$, at each time slot. Considering this action space leads our proposed MDP model to inherently capture the beamforming design. Let us proceed by defining the main components of the MDP model in our setup:
    i) Action Space: the action at time slot $i$ is denoted by 
$        \boldsymbol{\alpha}_i = [\alpha_{i, 1}, \ldots, \alpha_{i, K}, \alpha_{i, K+1}]$,
    where $\alpha_{i, K+1}$ is the action that determines the transmit power, such that $P_i^{\textrm{tx}} = \alpha_{i, K+1} P_{\max}$, and the action space is
        $\mathcal{A}_i = \bigl\{ \boldsymbol{\alpha}_i \in \mathbb{R}^{K + 1}| 
 \mathbf{0} \leq \boldsymbol{\alpha}_i \leq \mathbf{1} \bigr\}$;
    ii) State Space: the system state at time slot $i$ is determined by the available energy in the buffers and is given by
        $\mathbf{s}_i = [b_{i, 1}, \ldots, b_{i, K}]$
    and the state space can be written as
        $\mathcal{S}_i = \bigl\{\mathbf{s}_i \in \mathbb{R}^{K}| \mathbf{0} \leq \mathbf{s}_i \leq B_{\max}\mathbf{1}\bigr\}$;
    iii) Reward Function: the reward function should be able to capture \eqref{probnewa} and \eqref{probnewd}. Thus, the immediate reward at time slot $i$ is
    \begin{equation}\label{eq:rew_imd}
        {r}_{i} = 
             -\rho_1 e^{\bar{N}_i + \rho_2 \Delta_B} - e^{\frac{P_i^{\textrm{tx}}}{P_{\max}}},
    \end{equation}
    where $\rho_1 \geq 1$ and $0 \leq \rho_2 \leq 1$ are the weighting terms, 
    \begin{equation}
        \Delta_B = \frac{1}{B_{\max}}\sum_{k = 1}^K {(d_{i, k} +  P_o - b_{i,k})}\mathbbm{1}{(d_{i, k} +  P_o > b_{i,k})},
    \end{equation}
    is the normalized total deficiency of energy in the devices, and $\bar{N}_i = \sum_{k=1}^K \mathbbm{1}{\bigl(d_{i, k} +  P_o > b_{i,k}\bigr)}$ is the number of devices that were unable to meet their requirements at the $i$th slot. This reward indicates that the system is penalized based on the number of unsatisfied devices and their buffer state, and compels the system to reduce energy outage before minimizing the transmit power. Hereby, the discounted cumulative reward  at the end of the time slot $i$ is
        ${R}_i = \sum_{u = i}^{\infty} \gamma^{u - i} {r}_u$,
    where $0 \leq \gamma \leq 1$ is the discount factor, i.e., a lower $\gamma$ puts more emphasis on the immediate reward.


\subsection{Beamforming Design}

It is important to solve problem \eqref{probnewBF} given a specific action with low computation complexity since the system is time-sensitive. Let $\mathbf{w}^\star_{i, m}$ be the optimal solution of \eqref{probnewBF}. It has been proven \cite{proof1prob, proof2prob} that there exists $\mu_k^\star$ and $\beta_k^\star$ values such that $\mathbf{w}^\star_{i, m}$ is the optimal solution to the following problem:
\begin{align}
\label{probnewBFa} \argmax_{\mathbf{w}_{i, m} \in \mathcal{F}, \forall m} \quad &  \sum_{k = 1}^{K} \alpha_{i, k} \mu_k^\star \bigl[P^{\textrm{sat}}_k - \beta_k^\star(1 + e^{-\varphi_k({P^{\textrm{rf}}_{i,k}} - \omega_k)})\bigr],
\end{align}
where $\mathcal{F}$ is the set of feasible precoders satisfying \eqref{probnewBFb}, while ${P^{\textrm{rf}}_{i, k}} = \sum_{m = 1}^{M} ||\mathbf{a}^H_k \mathbf{w}_{i,m}||^2$ satisfies the following equations:
\begin{subequations}
\begin{align}
    \beta^\star_k (1 + e^{-\varphi_k({P^{\textrm{rf}}_{i,k}}^\star - \omega_k)}) - P^{\textrm{sat}}_k &= 0, \\
    \mu^\star_k (1 + e^{-\varphi_k({P^{\textrm{rf}}_{i,k}}^\star - \omega_k)}) - 1 &= 0.
    \vspace{-2mm}
\end{align}
\end{subequations}
This problem can be solved by an iterative algorithm including an inner loop for finding $\mathbf{w}_{i,m}$ given $\mu_k$ and $\beta_k$, and an outer loop for updating $\mu_k$ and $\beta_k$. One possible approach to solve the inner loop's problem is to reformulate the problem as a semi-definite program (SDP), as in \cite{sigmoidmodel}. However, this is not computationally/time efficient, especially for large $N$ since SDP deals with Hermitian matrix subspaces with size $N^2$.

\subsubsection{Suboptimal precoders given $\mu_k$ \& $\beta_k$}
 We consider $\mathbf{w}_{i, m}^{(l)}$ as the initial point and approximate \eqref{probnewBFa} by defining $P^{\textrm{rf}}_{i,k}$ as auxiliary variable and using the first-order Taylor expansion of the right-hand side of \eqref{eq:RCVPow} as
\begin{subequations}\label{probnewSCA}
\begin{align}
\label{probnewSCAa} \minimize_{{\mathbf{w}_{i,m} \in \mathcal{F}, P^{\textrm{rf}}_{i,k}}} \quad &  \Omega = - \sum_{k = 1}^{K} \alpha_{i, k} \mu_k^\star \bigl[P^{\textrm{sar}}_k - \beta_k^\star(1 + e^{-\varphi_k(P^{\textrm{rf}}_{i,k} - \omega_k)})\bigr]  \\
\textrm{subject to}  \label{probnewSCAc} \quad & P_{i,k}^{\textrm{rf}} \leq \sum_{m = 1}^M \biggl[2 \Re\bigl\{\mathbf{w}_{i, m}^H \mathbf{a}_k \mathbf{a}_k^H \mathbf{w}_{i, m}^{(l)}\bigr\} - \nonumber\\ &\hspace{25mm}{\mathbf{w}_{i, m}^{(l)}}^H \mathbf{a}_k \mathbf{a}_k^H \mathbf{w}_{i, m}^{(l)} \biggr],
\end{align}
\end{subequations}
which is convex in the neighborhood of the initial point. Then, the problem can be solved using successive convex approximation \cite{boyd2004convex}. Note that \eqref{probnewSCA} deals with complex vector subspace with size $N$, leading to less computational complexity in a single iteration compared to SDP. 
However, it still might challenge the system's real-time operation since the transmitter has to solve multiple subproblems in the inner loop. 

Let us proceed by writing the Lagrangian for \eqref{probnewSCA} as
\begin{multline}\label{eq:lagra}
    \mathcal{L}_i = - \sum_{k = 1}^{K} \alpha_{i, k} \mu_k^\star \bigl[P^{\textrm{sat}}_k - \beta_k^\star(1 + e^{-\varphi_k(P^{\textrm{rf}}_{i,k} - \omega_k)})\bigr] +  \\ \bar{\nu}_i(\sum_{m = 1}^M |\mathbf{w}_{i, m}|^2 - P_i^{\textrm{tx}}) +  \sum_{k = 1}^M \nu_{i,k} \biggl(P_{i,k}^{\textrm{rf}} - \\ \sum_{m = 1}^M \bigl[2 \Re\bigl\{\mathbf{w}_{i, m}^H \mathbf{a}_k \mathbf{a}_k^H \mathbf{w}_{i, m}^{(l)}\bigr\} - {\mathbf{w}_{i, m}^{(l)}}^H \mathbf{a}_k \mathbf{a}_k^H \mathbf{w}_{i, m}^{(l)} \bigr]\biggr),
\end{multline}
where $\nu_{i,k}$ and $\bar{\nu}_i$ are the associated dual variables. Then, using Karush–Kuhn–Tucker conditions \cite{boyd2004convex} and setting the derivative of \eqref{eq:lagra} w.r.t. $\mathbf{w}_{i,m}$ equal to zero, we can write
\begin{equation}\label{eq:beamformerstar}
     \mathbf{w}_{i,m}^\star = \frac{1}{\bar{\nu}^{(l - 1)}}\sum_{k = 1}^{K} \nu_{i,k} 
     \mathbf{a}_{i,k} \mathbf{a}_k^H \mathbf{w}_{i,m}^{(l-1)}.
\end{equation}
Notably, we introduce an additional parameter $\kappa$ to control the convergence of the proposed iterative method. Hereby, the precoders at the $l$th iteration are updated using $\mathbf{w}_{i,m}^{(l)} = \mathbf{w}_{i,m}^{(l-1)} + \kappa(\mathbf{w}_{i,m}^\star - \mathbf{w}_{i,m}^{(l - 1)})$. Obviously, the optimal solution of \eqref{probnewBF} utilizes the total amount of transmit power such that $\sum_{m = 1}^M ||\mathbf{w}_{i, m}||^2 = P^{\textrm{tx}}_i$, thus, $\bar{\nu}_i$ can be updated using
\begin{align}\label{eq:nubarup}
     \bar{\nu}^{(l)}_{i} = \sqrt{\frac{1}{P_i^{\textrm{tx}}}\sum_{m = 1}^{M} \bigl|\bigl|\sum_k \nu_{i,k} \mathbf{a}_k\mathbf{a_k}^H\mathbf{w}_{i,m}^{(l)}\bigr|\bigr|^2}.
\end{align}
Moreover, setting the derivative of \eqref{eq:lagra} w.r.t. $ P^{\textrm{rf}}_{i,k}$ equal to zero gives us 
\begin{align}\label{eq:nukup}
    \nu_{i,k}^{(l)} = \alpha_{i,k}\mu_k\beta_k \varphi_k e^{-\varphi_k({P_{i,k}^{\textrm{rf}}}^{(l)} - \omega_k)},
\end{align}
and ${P_{i,k}^{\textrm{rf}}}^{(l)}$ is obtained by substituting $\mathbf{w}_{i,m}^{(l)}$ in \eqref{eq:RCVPow}. 
These updating rules lead to low complexity since the beamformers can be derived using closed-form expressions.


\subsubsection{Updating $\mu_k$ \& $\beta_k$ given $\mathbf{w}_{i,m}$}

For this, we rely on the well-known damped Newton method \cite{proof2prob, sigmoidmodel}. Let us define $\phi_k  = \beta_k(1 + e^{-\varphi_k(P_{i,k}^{\textrm{rf}}-\omega_k)}) - P^{sat}_k$ and $\phi_{M + k} = \mu_k(1 + e^{-\varphi_k(P_{i,k}^{\textrm{rf}}-\omega_k)}) - 1$. It has been proven in \cite{proof1prob, proof2prob} that the unique optimal solution $\mu_k^\star$ and $\beta_k^\star$ is obtained if and only if $\boldsymbol{\phi}(\boldsymbol{\mu, \boldsymbol{\beta}}) = [\phi_1, \ldots, \phi_{2K}] = \mathbf{0}$, where $\boldsymbol{\mu} = [\mu_1, \ldots, \mu_K]$ and $\boldsymbol{\beta} = [\beta_1, \ldots, \beta_K]$. Therefore, $\mu_k$ and $\beta_k$ can be updated at the $n$th outer iteration using  
\begin{equation}\label{eq:mubetaupdate}
    \boldsymbol{\mu}^{(n + 1)}\!\!=\!\boldsymbol{\mu}^{(n)}\!\!+\!\zeta^{(n)}\mathbf{q}^{(n)}_{[K\!+\!1:2K]}, \ \boldsymbol{\beta}^{(n + 1)}\!\!=\!\boldsymbol{\beta}^{(n)}\!\!+\!\zeta^{(n)}\mathbf{q}_{[1:K]}^{(n)},
\end{equation}
where $\mathbf{q}_{[1:K]}^{(n)}$ refers to the first to $K$th element of $\mathbf{q}^{(n)}$, $\mathbf{q}^{(n)} = [\boldsymbol{\phi}'(\boldsymbol{\mu}, \boldsymbol{\beta})]^{-1} \boldsymbol{\phi}(\boldsymbol{\mu, \boldsymbol{\beta}})$ , $\boldsymbol{\phi}'(\boldsymbol{\mu}, \boldsymbol{\beta})$ is the Jacobian matrix of $\boldsymbol{\phi}(\boldsymbol{\mu}, \boldsymbol{\beta})$, and $\zeta^{(n)}$ is the largest $\epsilon^t$ satisfying 
\begin{equation}\label{eq:zetafind}
    ||\boldsymbol{\phi}(\boldsymbol{\mu}\!+\!\epsilon^t\mathbf{q}^{(n)}_{[K\!+\!1:2K]}, \boldsymbol{\beta}\!+\!\epsilon^t\mathbf{q}_{[1:K]}^{(n)})||\!\leq\!(1\!-\!\sigma\epsilon^t)||\boldsymbol{\phi}(\boldsymbol{\mu}, \boldsymbol{\beta})||,
\end{equation}
where $t \in \{1, 2, \ldots\}$, $\epsilon \in [0, 1]$, and $\sigma \in [0, 1]$.

\begin{algorithm}[t]
	\caption{\strut Iterative beamforming for the $i$th slot (IT-BF).} \label{alg:iterative}
	\begin{algorithmic}[1]
            \State \textbf{Input:}$\{\alpha_{i, m}\}_{\forall m}, P_i^{\textrm{tx}}$, $\epsilon$, $\zeta$, $\sigma$, $\Phi$, $\rho$
            \textbf{Output:} $\mathbf{w}_{i,m}^{(l)}, \forall m$
            \State \textbf{Initialize:} $\mathbf{w}^{(l)}_{i,m}, \forall m$,  $\bar{\nu}$, $\nu_k, \forall k$,  $\xi = 0$
            \Repeat\ (outer loop)\label{ALG:outerSTART}
                \hspace{-2mm}\Repeat\ (inner loop) \label{ALG:INNERSTART}
                    \State\hspace{-2mm} $\xi^\star \leftarrow \xi$
                    \State\hspace{-2mm} Obtain $\mathbf{w}_{i, m}^\star, \forall m$ using \eqref{eq:beamformerstar}
                    \State\hspace{-2mm}$\mathbf{w}_{i,m}^{(l)} \leftarrow \mathbf{w}_{i,m}^{(l)} + \rho(\mathbf{w}_{i,m}^\star - \mathbf{w}_{i,m}^{(l)}), \forall m$
                    \State \hspace{-2mm}Update $\bar{\nu}$ and $\nu_k, \forall k$ using \eqref{eq:nubarup} and \eqref{eq:nukup}
                    \State \hspace{-3mm}${P_{i,k}^{\textrm{rf}}}^{(l)} = \sum_{m = 1}^{M} ||\mathbf{a}^H_k \mathbf{w}^{(l)}_{i,m}||^2$, compute  $\xi$ using \eqref{probnewSCAa}
                \hspace{-2mm}\Until{$|1 - \frac{\xi}{\xi^\star}| < \Phi$}\label{ALG:INNERfinish}
                \State \hspace{-2mm} Obtain $\zeta^{(n)}$, update $\boldsymbol{\mu}^{(n)}$ and $\boldsymbol{\beta}^{(n)}$ using \eqref{eq:zetafind} and \eqref{eq:mubetaupdate}
            \Until{convergence}\label{ALG:outerfinish}
\end{algorithmic} 
\end{algorithm}

Algorithm~\ref{alg:iterative} illustrates the proposed beamforming approach at time slot $i$. Therein, the derived closed-form solutions are used to update the beamformers and the dual variables in the inner loop (lines \ref{ALG:INNERSTART}-\ref{ALG:INNERfinish}) until \eqref{probnewSCAa} converges, while $\mu_k$ and $\beta_k$ are updated in the outer loop (lines \ref{ALG:outerSTART}-\ref{ALG:outerfinish}). This procedure is repeated until a convergence criterion is satisfied.

\subsection{DDPG-based Joint Beamforming \& Charging Scheduling}


The goal is to maximize the long-term commutative reward by training a policy. For this, we utilize DDPG \cite{lillicrap2019continuous}, a model-free off-policy algorithm for learning continuous actions, as in our case. Let us denote the Q-value function as the expected commutative reward given by $\mathcal{Q}(\mathbf{s}_i, \boldsymbol{\alpha}_i) = \mathbb{E}\bigl\{{R}_i\bigr\}$, while the target Q-network is $\mathcal{Q}^\star$. DDPG approximates the policy and Q-value functions utilizing two neural networks (NNs), defined as the actor and the critic (see details on the operation of the actor-critic approach in \cite{lillicrap2019continuous}). The following outlines our DDPG implementation of the formulated MPD. 


Let us define the parameterized actor function $\mu$ as the current policy by deterministically mapping states to a specific action. By defining $\theta^\mu$ and $\theta^\mathcal{Q}$ as the parameters of the actor and critic NNs, respectively, the actor NN parameters are updated by applying the chain rule to the expected return from the start distribution ($J$) such that \cite{lillicrap2019continuous}
\begin{align}\label{eq:DPGtheorem}
     \nabla_{\!\theta^\mu}\! J\!\approx\!\mathbb{E}\biggl\{\!\!\nabla_{\boldsymbol{\alpha}}\mathcal{Q}(\mathbf{s}, \boldsymbol{\alpha}|{\theta^Q})\bigl|_{\mathbf{s}\!=\!\mathbf{s}_i, \boldsymbol{\alpha}\!=\!\mu(\mathbf{s}_i)} \!\!\!\nabla_{\!\theta^\mu}\mu(\mathbf{s}|\theta^\mu)|_{\mathbf{s} = \mathbf{s}_i} \!\!\biggr\}.
\end{align}
Meanwhile, the critic NN aims to learn $\mathcal{Q}(\mathbf{s}_i, \alpha_i)$. It has been proven that each Q-function for some policy obeys the Bellman equation, thus, we write \cite{DRL-survey}
\begin{equation}
    y_i = {r}_i + \gamma \mathcal{Q}^\star\bigl(s_{i+ 1}, \mu^\star(s_{i + 1}|\theta^{\mu^\star})|\theta^{Q^\star}\bigr),
\end{equation}
where $\mu^\star$ is the target actor. The critic NN aims to minimize a loss function of the temporal difference error given by
\begin{equation}\label{eq:loss}
    \mathcal{L}_i = \frac{1}{|\mathcal{B}_i|} \sum_{(\mathbf{s}_i, \boldsymbol{\alpha}_{i}, \mathbf{s}_{i + 1}, {r}_{i}) \in \mathcal{B}_i} [\mathcal{Q}(\mathbf{s}_i, \boldsymbol{\alpha}_i) - y_i]^2,
\end{equation}
where $\mathcal{B}_i \subseteq \mathcal{B}$ is the minibatch of transitions and $\mathcal{B}$ is the memory buffer. Notice that the system includes a replay memory to store the transitions for future use.

To make the system observable, the active devices send a cost-free reliable\footnote{In practice, the feedback comes with an associated transmission error and cost depending on the modulation and the message length. However, for simplicity and without loss of generality, we assume successful message delivery and that the devices always save energy for feedback.} feedback message at the end of each time slot declaring their past required energy and whether they had sufficient energy to meet that. If the transmitter does not receive any feedback, it assumes that the device is in idle mode. Hereby, the system becomes observable and the transmitter can update the states after each time slot. 

Algorithm~\ref{alg:DQL} illustrates the proposed DDPG-based joint beamforming and charging scheduling. At first, the system state and a random process $\mathcal{N}$ for the exploration noise are initialized. Each episode consists of $T$ time slots and at the beginning of each time slot, an action is selected using the actor NN and a generated exploration noise. This noise leads to a trade-off between exploration and exploration, leading to better convergence. Then, Algorithm~\ref{alg:iterative} is run to obtain the beamformers for the selected action. Finally, the actor and critic NNs are updated, and the corresponding target NNs' parameters are softly updated using $\tau$, which leads to small changes in target NNs; thus, offering stability in training \cite{lillicrap2019continuous}.

\begin{algorithm}[t]
	\caption{\strut DDPG-based joint beamforming and charging scheduling (DDPG-BCS).} \label{alg:DQL}
	\begin{algorithmic}[1]
            \State \textbf{Input:} $\gamma$, $\theta^\mu$, $\theta^{\mu^\star}$, $\theta^\mathcal{Q}$, $N_{ep}$, $\theta^{\mathcal{Q}^\star}$, $\tau$
             \textbf{Output:} $\mu^\star$ 
            \State Get the initial system state $\mathbf{s}_1$
            \For{$q = 1, 2, \ldots, N_{ep}$}\label{algDDPG:epstart}
                \State Initialize a random process $\mathcal{N}$ for exploration
                \For{$i = 1, 2, \ldots, T$}\label{algDDPG:slotstart} 
                    \State \hspace{-2mm} Generate a random noise $\mathbf{n}$ using  $\mathcal{N}$
                    \State \hspace{-2mm} Select an action using $\boldsymbol{\alpha}_i = \mu(\mathbf{s}_i|\theta^\mu) + \mathbf{n}$ 
                    \State \hspace{-2mm} Run IT-BF to obtain $\mathbf{w}_{i,m}$ 
                    \State \hspace{-2mm} Compute $r_i$ and $\mathbf{s}_{i + 1}$ using \eqref{eq:bufermodel}, and \eqref{eq:rew_imd}
                    \State \hspace{-2mm} Store the transition $(\mathbf{s}_{i}, \boldsymbol{\alpha}_{i}, \mathbf{s}_{i + 1}, {r}_{i})$ in $\mathcal{B}$
                    \State \hspace{-2mm} Sample a random minibatch $\mathcal{B}_i \subseteq \mathcal{B}$ with size $N_b$ 
                    \State \hspace{-2mm} Update the critic NN by minimizing $\mathcal{L}_i$ in \eqref{eq:loss}
                    \State \hspace{-2mm} Update the actor NN by sampling \eqref{eq:DPGtheorem} for $\mathcal{B}_i$
                    \State \hspace{-2mm} $\theta^{\mu^\star} \leftarrow \tau \theta^{\mu} + (1-\tau)\theta^{\mu^\star}$, $\theta^{\mathcal{Q}^\star} \leftarrow \tau \theta^{\mathcal{Q}} + (1-\tau)\theta^{\mathcal{Q}^\star}$
                \EndFor\label{algDDPG:slotend} 
                \State $\mathbf{s}_1 \leftarrow \mathbf{s}_{T}$
            \EndFor\label{algDDPG:epend} 
\end{algorithmic} 
\end{algorithm}

\section{Numerical Analysis}\label{SEC:numerical}

We consider a square uniform planar array with $N = 64$, inter-element distance $\frac{\lambda}{2}$, and $\lambda = 12.5$ cm, corresponding to operation at 2.4 GHz. The transmitter is at the ceiling's center of a $5\times 5 \times 5$~m$^3$ area, while the users are uniformly positioned on a circle along the $xy$ dimension with a 2~m radius such that the position of $k$th device is $[2.5 + 2\cos{(2\pi/K)}, 2.5 + 2\sin(2\pi/K), 2]$. Moreover, $\mathcal{N}$ follows a Ornstein-Uhlenbeck process with parameters $\theta^{\mathcal{N}} = 0.15$, $\sigma^\mathcal{N} = 0.2$ \cite{lillicrap2019continuous}. For device activation, we utilize an alarm generation process based on the spatial location of the devices, in which an alarm activates a set of devices based on the activation probability function $f(\chi_k) = e^{-\chi_k}$ with $\chi_k$ being the distance of the $k$th device from the event epicenter. Notably, the epicenter is generated using a uniform distribution across the area. Moreover, when a device is activated, the required amount of energy is selected using a geometric distribution with parameter $\vartheta = 0.5$. Thus, there is an energy demand burst  $nd_{b}$ for an activated device with probability $\vartheta^{n - 1}(1 - \vartheta)$, while the maximum possible demand is 50~mW, $d_{b} = 10$~mW, and $P_o = 10$~$\mu$W \cite{3gppamb}. The initial battery level of the devices is 50~mW, $B = 200~mW$, and $P_{\max} = 10$~W. The EH device parameters are $\phi_k = 6400$, $\omega_k = 0.003$, $P_k^{\textrm{
sat}} = 20$~mW \cite{sigmoidmodel}.

The actor NN consists of two fully connected layers with 128 neurons and a Rectified Linear Unit (ReLU) activation function. Moreover, the output activation function is tangent hyperbolic, leading to the range $[-1, 1]$, thus, the outputs are scaled to fit in the desired range $[0, 1]$. The critic NN has two hidden layers with 64 neurons and ReLU activation function, while the output utilizes linear activation. The IT-BF algorithm convergence criterion is reaching 20 outer loop iterations or $\boldsymbol{\phi}(\boldsymbol{\mu}, \boldsymbol{\beta}) \leq 10^{-4} \mathbf{1}$. The optimization parameters are $T = 100$, $\kappa = 0.05$, $g = 2$, $\tau = 0.001$, $N_b = 64$, $|B| = 10^6$, $\gamma = 0.99$, $\Phi = 10^{-6}$, $\rho_1 = \rho_2 = 1$, $\sigma = 0.5$, $\epsilon = 0.5$, while actor and critic learning rates are $10^{-4}$ and $2\times10^{-4}$, respectively. We compare the performance of our proposed DDPG-BCS approach with a heuristic aiming to keep the buffer state of the devices above the maximum energy demand (50 mW). Thus, at the $i$th time slot, the set of devices with the buffer value below 50 mW, denoted as $\mathcal{K}_i$, is charged with $P_i^{\textrm{tx}} = P_{\max}$, $\alpha_{i,k} = \bar{b}_{i,k}/\sum_{k \in \mathcal{K}_i}\bar{b}_{i,k}$, and $\bar{b}_{i,k} = 50 - b_{i,k}$.

\begin{figure}[t]
    \centering
    \includegraphics[width=0.75\columnwidth]{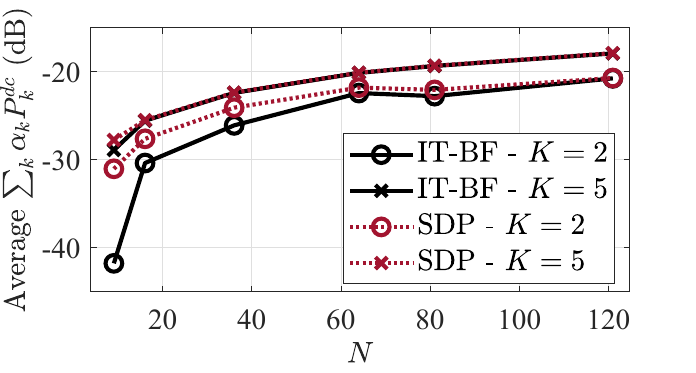}
    \vspace{-3mm}
    \includegraphics[width=0.75\columnwidth]{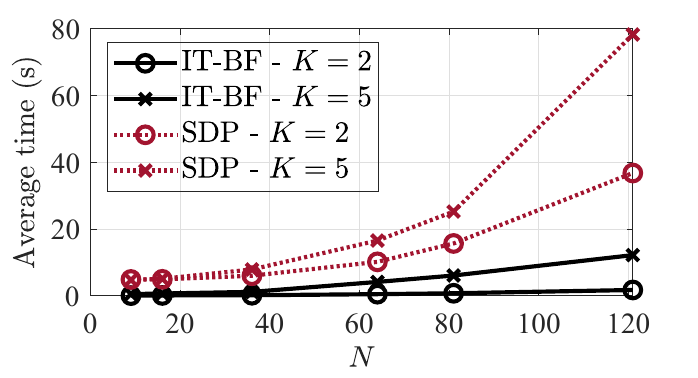}
    \caption{(a) The average sum harvested power (top) and (b) the convergence time (bottom) as a function of $N$.}
    \label{fig:iterBF}
    \vspace{-4mm}
\end{figure}

\begin{figure}[t]
    \centering
    \includegraphics[width=0.8\columnwidth]{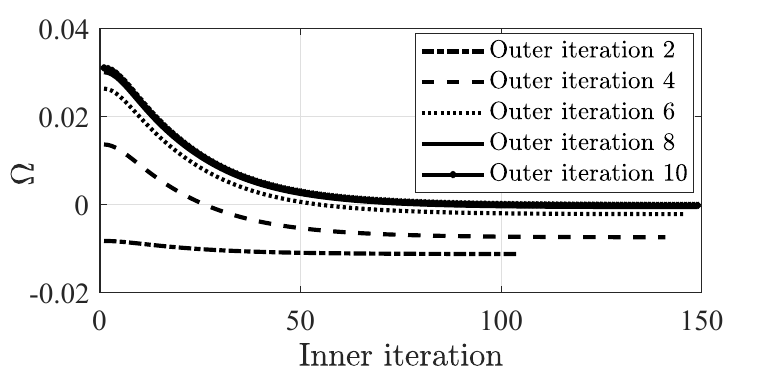}
    \caption{The $\Omega$ value as a function of the inner loop iterations for $K = 6$, $P^{\textrm{tx}} = 10$~W, and random $\alpha_{k}$.}
    \label{fig:converge}
    \vspace{-6mm}
\end{figure}

Fig.~\ref{fig:iterBF} illustrates the average performance of the proposed IT-BF using Monte Carlo simulations. It is seen that the IT-BF performs near the optimal SDP solution and the bound becomes tighter as $N$ increases. Meanwhile, IT-BF requires considerably less time \footnote{Note that the convergence time is obtained using an average-performance computer, and the purpose is to only show the relative time difference. However, the convergence time can become significantly lower by using proper hardware resources.} to converge and the complexity increases with $N$ and $K$. Note that the convergence performance of IT-BF, presented in Fig.~\ref{fig:converge}, shows that the algorithm converges toward a suboptimal solution with outer iterations.

\begin{figure}[t]
    \centering
    \includegraphics[width=0.85\columnwidth]{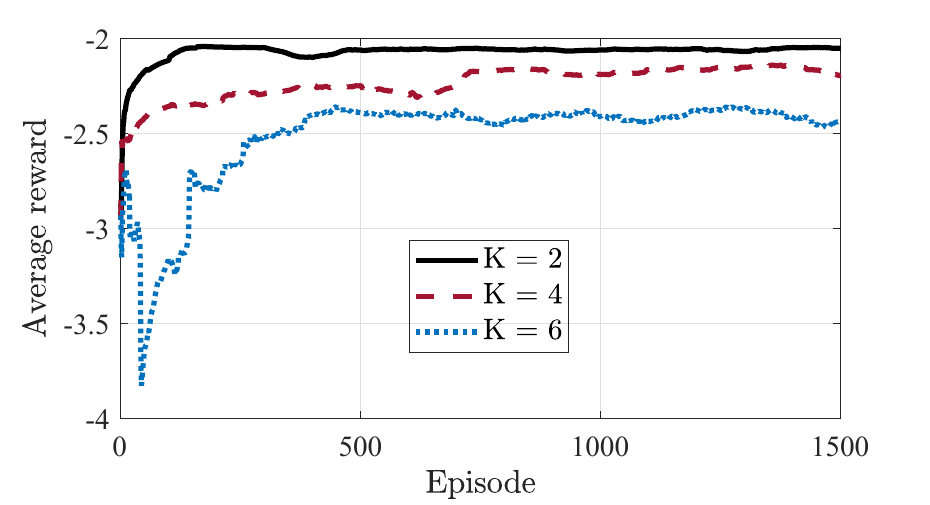}
    \caption{The average reward as a function of the number of episodes for different numbers of devices.}
    \label{fig:reward}
    \vspace{-6mm}
\end{figure}

\begin{figure}[t]
    \centering
    \includegraphics[width=0.85\columnwidth]{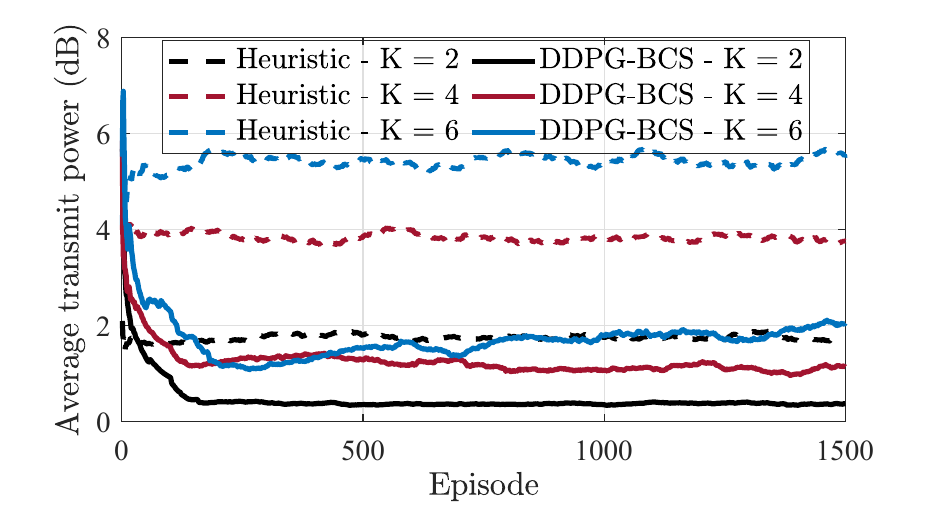}
    \includegraphics[width=0.85\columnwidth]{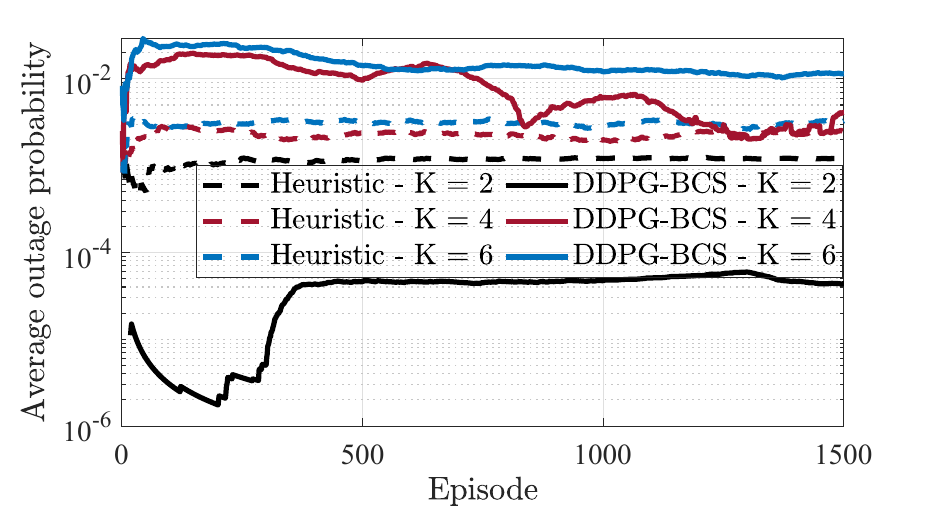}
    \caption{(a) The average transmit power (top), (b) the average outage probability (bottom) as a function of the number of episodes for different numbers of devices.}
    \label{fig:powoutage}
\end{figure}

Fig.~\ref{fig:reward} showcases the reward function of DDPG-BCS over learning episodes. It is seen that the reward gradually converges toward a suboptimal solution for each $K$. Meanwhile, Fig.~\ref{fig:powoutage} shows the average outage probability, i.e., $(\sum_{i = 1}^T \bar{N}_i)/KT$, and the average transmit power over episodes. Observe that the DDPG-BCS outperforms the heuristic in terms of power consumption with a considerable gap. However, as the number of devices increases, the heuristic leads to fewer outages, but with much higher transmit power. Instead, the proposed approach aims for the trade-off between outage and transmit power, which can be modified by shaping the reward function. In general, it is seen that the power consumption increases with $K$, while more outages may happen as $K$ becomes larger. Notably, even for $K = 6$ the outage probability is relatively small ($\leq 10^{-2}$), while for lower $K$, this value is near zero.

\vspace{-1mm}

\section{Conclusion and Future Work }\label{SEC:conclusion}

\vspace{-1mm}

In this paper, we considered a multi-antenna RF-WPT system to charge multiple non-linear EH devices over time. Moreover, we formulated the joint beamforming and charging scheduling problem to minimize the average transmit power and meet the EH requirements and proposed a solution relying on DDPG and a low-complexity beamforming design. The results demonstrated that our proposed beamforming approach achieves a near-optimal solution and that the DDPG-based optimization converges after some episodes. It was shown that the outage probability and the energy consumption increase and decrease with the number of devices, respectively.

As a prospect for future research, one may investigate the system performance in a partially observable environment or consider channel variations due to device mobility.

\vspace{-2mm}
\bibliographystyle{ieeetr}
\bibliography{ref}


%




\end{document}